\documentclass[aps,prd,superscriptaddress,nofootinbib,tighten,preprint]{revtex4}
\usepackage{amsfonts}
\usepackage{amssymb}
\usepackage{amsmath,color}
\usepackage{epsfig}
\usepackage{graphicx,epsf,epsfig,subfigure}
\usepackage{bbm}
\def\slc#1{\setbox0=\hbox{$#1$}           
    \dimen0=\wd0                                 
    \setbox1=\hbox{/} \dimen1=\wd1               
    \ifdim\dimen0>\dimen1                        
       \rlap{\hbox to \dimen0{\hfil/\hfil}}      
       #1                                        
    \else                                        
       \rlap{\hbox to \dimen1{\hfil$#1$\hfil}}   
       /                                         
    \fi}

\begin{document}

\title{Extrinsic and Intrinsic CPT Asymmetries in Neutrino Oscillations}

\author{Tommy Ohlsson}

\email{tohlsson@kth.se}

\affiliation{Department of Theoretical Physics, School of
Engineering Sciences, KTH Royal Institute of Technology, AlbaNova
University Center, 106 91 Stockholm, Sweden}

\author{Shun Zhou}
\email{shunzhou@kth.se}

\affiliation{Department of Theoretical Physics, School of
Engineering Sciences, KTH Royal Institute of Technology, AlbaNova
University Center, 106 91 Stockholm, Sweden}

\affiliation{Institute of High Energy Physics, Chinese Academy of Sciences, Beijing 100049, China}

\begin{abstract}
We reconsider the extrinsic and possible intrinsic CPT violation in neutrino oscillations, and point out an identity, i.e., $A^{\rm CP}_{\alpha \beta} = A^{\rm CPT}_{\beta \alpha} + A^{\rm T}_{\alpha \beta}$, among the CP, T, and CPT asymmetries in oscillations. For three-flavor oscillations in matter of constant density, the extrinsic CPT asymmetries $A^{\rm CPT}_{ee}$, $A^{\rm CPT}_{e\mu}$, $A^{\rm CPT}_{\mu e}$, and $A^{\rm CPT}_{\mu \mu}$ caused by Earth matter effects have been calculated in the plane of different neutrino energies and baseline lengths. It is found that two analytical conditions can be implemented to describe the main structure of the contours of vanishing extrinsic CPT asymmetries. Finally, without assuming intrinsic CPT symmetry in the neutrino sector, we investigate the possibility to constrain the difference of the neutrino CP-violating phase $\delta$ and the antineutrino one $\overline{\delta}$ using a low-energy neutrino factory and the super-beam experiment ESS$\nu$SB.  We find that $|\delta - \overline{\delta}| \lesssim 0.35\pi$ in the former case and $|\delta - \overline{\delta}| \lesssim 0.7\pi$ in the latter case can be achieved at the $3\sigma$ confidence level if $\delta = \overline{\delta} = \pi/2$ is assumed.
\end{abstract}

\maketitle

\section{introduction}
\label{sec:intro}

Recent years have seen great progress in experimental neutrino physics. In particular, neutrino oscillations have been well established and leptonic mixing parameters have been measured with an acceptable degree of accuracy. Under the assumption of conservation of the fundamental CPT symmetry, both three-flavor neutrino and antineutrino oscillations can be described by the same set of parameters, namely three leptonic mixing angles $(\theta_{12}, \theta_{13}, \theta_{23})$, one leptonic Dirac CP-violating phase $\delta$, and two independent mass-squared differences $(\Delta m^2_{21}, \Delta m^2_{31})$, where $\Delta m^2_{21} \equiv m^2_2 - m^2_1$ and $\Delta m^2_{31} \equiv m^2_3 - m^2_1$ with $(m_1, m_2, m_3)$ being the three neutrino masses. The primary goals of present and future neutrino oscillation experiments are to perform precision measurements of the neutrino parameters, determine the neutrino mass ordering (i.e., the sign of $\Delta m^2_{31}$), and probe $\delta$. In the future, one could also try to establish if there is fundamental or intrinsic CPT violation in the neutrino sector.

Previously, various theoretical models based on violation of the fundamental CPT symmetry have been proposed in the literature. Such models, that naturally also break Lorentz invariance \cite{Greenberg:2002uu}, include works by Coleman \& Glashow \cite{Coleman:1997xq,Coleman:1998ti} and Kosteleck{\'y} et al.~\cite{Kostelecky:2003xn,Kostelecky:2003cr,Diaz:2009qk,Diaz:2011ia}. On the more phenomenological side, studies of CPT violation have recently been performed in Refs.~\cite{Barger:2000iv,Murayama:2000hm,Barenboim:2001ac,Bilenky:2001ka,Barenboim:2002rv,Bahcall:2002ia,Barenboim:2002hx,Datta:2003dg,Minakata:2005jy,Dighe:2008bu,Antusch:2008zj,Barenboim:2009ts,Samanta:2010ce,Giunti:2010zs,Chatterjee:2014oda}. Indirect limits on CPT violation for specific models in the neutrino sector have also been presented \cite{Mocioiu:2002pz}. Finally, experimental collaborations have searched for signals of CPT violation in neutrino oscillation experiments, which include LSND \cite{Auerbach:2005tq}, MiniBooNE \cite{AguilarArevalo:2011yi,Katori:2014tpa}, MINOS \cite{Adamson:2008aa,Adamson:2010rn,Adamson:2012hp,Cao:2014eca}, and Super-Kamiokande \cite{Abe:2011ph,Kaji:2011zz,Takeuchi:2011aa}.

In a phenomenological way, if the CPT symmetry is not assumed a priori, we need two separate sets of parameters to describe neutrino and antineutrino oscillations. Now, the neutrino flavor eigenstates $|\nu_\alpha \rangle $ are related to the neutrino mass eigenstates $|\nu_i\rangle$ by a $3 \times 3$ unitary leptonic mixing matrix (see e.g.~Ref.~\cite{Bilenky:1987ty})
\begin{equation}
|\nu_\alpha \rangle = \sum^3_{i=1} U^*_{\alpha i}(\theta_{12}, \theta_{13}, \theta_{23}, \delta) |\nu_i\rangle
\label{eq:nu}
\end{equation}
and the three neutrino masses are $m_i$ (for $i = 1, 2, 3$). Similarly, for antineutrinos, we have
\begin{equation}
|\overline{\nu}_\alpha \rangle = \sum^3_{i=1} U_{\alpha i}(\overline{\theta}_{12}, \overline{\theta}_{13}, \overline{\theta}_{23}, \overline{\delta}) |\overline{\nu}_i\rangle
\label{eq:nubar}
\end{equation}
and the antineutrino masses are denoted by $\overline{m}_i$ (for $i = 1, 2, 3$). Therefore, the mass-squared differences of antineutrinos are defined as $\Delta \overline{m}^2_{21} \equiv \overline{m}^2_2 - \overline{m}^2_1$ and $\Delta \overline{m}^2_{31} \equiv \overline{m}^2_3 - \overline{m}^2_1$. Although micro-causality may be violated if the masses of particles and the masses of their corresponding antiparticles are different from each other, the results in our phenomenological approach can actually be applied to the scenario of spontaneous CPT violation in Refs.~\cite{Coleman:1997xq, Coleman:1998ti, Kostelecky:2003xn, Kostelecky:2003cr}. In principle, neutrino oscillation experiments can be used to place restrictive constraints on the CPT-violating parameters in the neutrino sector.

However, in long-baseline neutrino oscillation experiments, such as a future neutrino factory, neutrinos and antineutrinos will traverse Earth matter, and therefore, matter effects on neutrino and antineutrino oscillations will induce fake or extrinsic CPT-violating effects.

In this work, we investigate the extrinsic and intrinsic CPT asymmetries in oscillations. First, some general remarks are given on the relationship among the CP, T, and CPT asymmetries. An identity $A^{\rm CPT}_{\beta \alpha} + A^{\rm T}_{\alpha \beta} = A^{\rm CP}_{\alpha \beta}$ is derived. Second, we explore the conditions under which the extrinsic CPT asymmetries induced by matter effects vanish. In this case, if intrinsic CPT violation exists, it will be made more apparent. Finally, we illustrate the experimental sensitivity to the CPT-violating parameters by taking a low-energy neutrino factory and a super-beam experiment as examples.

\section{CP, T, and CPT Asymmetries}
\label{sec:cpt}

First of all, we present some general discussion on the CP, T, and CPT asymmetries in neutrino and antineutrino oscillations in vacuum and matter (see e.g.~Ref.~\cite{Jacobson:2003wc} and references therein). We denote the oscillation probabilities for neutrinos in the $\nu^{}_\alpha \to \nu^{}_\beta$ channels by $P^{}_{\alpha \beta} = P(\nu^{}_\alpha \to \nu^{}_\beta)$, while those for antineutrinos by $\overline{P}^{}_{\alpha \beta} = P(\overline{\nu}^{}_\alpha \to \overline{\nu}^{}_\beta)$. Here the neutrino flavor indices $\alpha$ and $\beta$ run over $e$, $\mu$, and $\tau$. Note that the oscillation probabilities $P^{}_{\alpha \beta}$ are dependent due to the unitarity conditions:
\begin{eqnarray}
\sum_\alpha P^{}_{\alpha \beta} = \sum_\beta P^{}_{\alpha \beta} = 1 \; ,
\label{eq:unitarity}
\end{eqnarray}
and likewise for $\overline{P}^{}_{\alpha \beta}$.
It is straightforward to verify that four out of nine oscillation probabilities in the three-flavor case are independent \cite{Jacobson:2003wc}. However, in the two-flavor case, there is only one independent oscillation probability. Based on neutrino and antineutrino oscillation probabilities, the CP, T, and CPT asymmetries can be defined as
\begin{eqnarray}
A^{\rm CP}_{\alpha \beta} &\equiv& P^{}_{\alpha \beta} - \overline{P}^{}_{\alpha \beta} \; , \label{eq:CP}\\
A^{\rm T}_{\alpha \beta} &\equiv& P^{}_{\alpha \beta} - P^{}_{\beta \alpha} \; , \label{eq:T}\\
A^{\rm CPT}_{\alpha \beta} &\equiv& P^{}_{\alpha \beta} - \overline{P}^{}_{\beta \alpha} \; . \label{eq:CPT}
\end{eqnarray}
Hence, any CP, T, and CPT violation will be characterized by a non-zero value of $A^{\rm CP}_{\alpha \beta}$, $A^{\rm T}_{\alpha \beta}$, and $A^{\rm CPT}_{\alpha \beta}$, respectively. In a similar way, one can also define the corresponding asymmetries for antineutrinos, i.e., $\overline{A}^{\rm CP}_{\alpha \beta}$, $\overline{A}^{\rm T}_{\alpha \beta}$, and $\overline{A}^{\rm CPT}_{\alpha \beta}$. Obviously, $\overline{A}^{\rm CP}_{\alpha \beta}$ and $\overline{A}^{\rm CPT}_{\alpha \beta}$ are dependent quantities, since they are related to the CP and CPT asymmetries for neutrinos, i.e., $\overline{A}^{\rm CP}_{\alpha \beta} = - A^{\rm CP}_{\alpha \beta}$ and $\overline{A}^{\rm CPT}_{\alpha \beta} = - A^{\rm CPT}_{\beta \alpha}$. However, the T asymmetries $\overline{A}^{\rm T}_{\alpha \beta} = \overline{P}^{}_{\alpha \beta} - \overline{P}^{}_{\beta \alpha}$ are in general independent.

Subtracting Eq.~(\ref{eq:CP}) from Eq.~(\ref{eq:CPT}), one obtains an interesting relation among the CP, T, and CPT asymmetries, viz.
\begin{eqnarray}
A^{\rm CP}_{\alpha \beta} + \overline{A}^{\rm T}_{\alpha \beta} = A^{\rm CPT}_{\alpha \beta} \; , \label{eq:idnu}
\end{eqnarray}
and similarly, for the antineutrino counterpart of Eqs.~(\ref{eq:CP}) and (\ref{eq:CPT}), we find that
\begin{eqnarray}
\overline{A}^{\rm CP}_{\alpha \beta} + A^{\rm T}_{\alpha \beta} = \overline{A}^{\rm CPT}_{\alpha \beta} \; . \label{eq:idnubar}
\end{eqnarray}
Now, it is straightforward to derive
\begin{eqnarray}
A^{\rm CPT}_{\beta \alpha} + A^{\rm T}_{\alpha \beta} = A^{\rm CP}_{\alpha \beta} \; , \label{eq:id}
\end{eqnarray}
and a similar relation among the corresponding asymmetries for antineutrinos. It is worthwhile to emphasize that the relation in Eq.~(\ref{eq:id}) is valid even if the fundamental CPT symmetry is not preserved. Some comments are in order:
\begin{itemize}
\item From the definition in Eq.~(\ref{eq:T}), we can observe that $A^{\rm T}_{\alpha \beta} = -A^{\rm T}_{\beta \alpha}$, and thus, the T asymmetry $A^{\rm T}_{\alpha \beta}$ vanishes in the disappearance channels $\alpha = \beta$, i.e., $A^{\rm T}_{ee} = A^{\rm T}_{\mu \mu} = A^{\rm T}_{\tau \tau} = 0$ \cite{Krastev:1988yu,Akhmedov:2001kd}. Furthermore, Eq.~(\ref{eq:unitarity}) implies $\sum_\alpha A^{\rm T}_{\alpha \beta} = \sum_\beta A^{\rm T}_{\alpha \beta} = 0$, so we have $A^{\rm T}_{e\mu} = A^{\rm T}_{\mu\tau} = A^{\rm T}_{\tau e} \equiv {\cal A}^{\rm T}$, which is the unique T asymmetry in the three-flavor case  \cite{Krastev:1988yu}. This conclusion applies to oscillations both in vacuum and matter.

\item For oscillations in vacuum, there is no extrinsic CPT violation \cite{Jacobson:2003wc}. Then, if the intrinsic CPT symmetry holds, we can see that $A^{\rm T}_{\alpha \beta} = A^{\rm CP}_{\alpha \beta}$ from Eq.~(\ref{eq:id}), implying a unique CP asymmetry ${\cal A}^{\rm CP}$. Explicitly, we can calculate the unique CP or T asymmetry \cite{Krastev:1988yu}
    \begin{equation}
    {\cal A}^{\rm T} = {\cal A}^{\rm CP} = 16 {\cal J} \sin \frac{\Delta m^2_{21}L}{4E} \sin \frac{\Delta m^2_{32}L}{4E} \sin \frac{\Delta m^2_{31}L}{4E} \; ,
    \label{eq:unique}
    \end{equation}
    where ${\cal J} \equiv \sin \theta_{12} \cos \theta_{12} \sin \theta_{23} \cos \theta_{23} \sin \theta_{13} \cos^2 \theta_{13} \sin \delta$ is the Jarlskog invariant \cite{Jarlskog:1985cw,Jarlskog:1985ht} and $\Delta m^2_{32} \equiv m^2_3 - m^2_2$. Here $E$ is the neutrino beam energy and $L$ is the baseline length. However, in the presence of matter effects or intrinsic CPT violation, we have four independent CP asymmetries $A^{\rm CP}_{\alpha \beta}$, as indicated by $\sum_\alpha A^{\rm CP}_{\alpha \beta} = \sum_\beta A^{\rm CP}_{\alpha \beta} = 0$ and $A^{\rm CP}_{\alpha \alpha} \neq 0$. This applies also to the CPT asymmetries $A^{\rm CPT}_{\alpha \beta}$.
\end{itemize}

In the following section, we will calculate the CPT asymmetries for oscillations in matter, assuming constant matter density. Furthermore, the conditions, under which the extrinsic CPT asymmetries vanish, will be derived and discussed.

\section{Extrinsic CPT Asymmetries}
\label{sec:ext}

We proceed to consider CPT asymmetries in two- and three-flavor neutrino and antineutrino oscillations in matter. For oscillations in vacuum, the CPT asymmetries $A^{\rm CPT}_{\alpha \beta}$ vanish exactly if the fundamental CPT symmetry is preserved. It has been pointed out that one can test the intrinsic CPT symmetry in a long-baseline neutrino oscillation experiment, such as a future neutrino factory \cite{Bilenky:2001ka}. As we will show in the next section, future neutrino superbeam experiments and low-energy neutrino factories are very powerful in constraining intrinsic CPT violation, particularly for the case of CP-violating phases. However, the extrinsic CPT asymmetry induced by Earth matter effects will mimick the intrinsic one, reducing experimental sensitivity to the CPT-violating parameters. Therefore, it is interesting to explore the conditions for the extrinsic CPT asymmetries to vanish.

\subsection{Two-Flavor Case}

For two-flavor oscillations, say $\nu^{}_e$ and $\nu^{}_\mu$, only one probability for neutrinos is independent, and we choose $P^{}_{ee}$. Similarly, we take $\overline{P}^{}_{ee}$ for antineutrinos. In this case, we have $P^{}_{e\mu} = 1 - P^{}_{ee} = P^{}_{\mu e}$, so the T asymmetries are $A^{\rm T}_{e \mu} \equiv P^{}_{e\mu} - P^{}_{\mu e} = 0$ and $A^{\rm T}_{\mu e} \equiv P^{}_{\mu e} - P^{}_{e \mu} = 0 $. Note that $A^{\rm T}_{ee} = A^{\rm T}_{\mu \mu} = 0$ by definition. However, there is only one CPT asymmetry
\begin{equation}
A^{\rm CPT}_{e \mu} = P^{}_{e \mu} - \overline{P}^{}_{\mu e} = (1 - P^{}_{ee}) - (1 - \overline{P}^{}_{ee}) = \overline{P}^{}_{ee}  - P^{}_{ee} = -A^{\rm CPT}_{ee} \; .
\end{equation}
One can further verify that $A^{\rm CPT}_{ee} = A^{\rm CPT}_{\mu \mu} = - A^{\rm CPT}_{e\mu} = - A^{\rm CPT}_{\mu e} \equiv A^{\rm CPT}_{2\nu}$. According to Eq.~(\ref{eq:id}), we find that $A^{\rm CP}_{ee} = A^{\rm CP}_{\mu \mu} = - A^{\rm CP}_{e\mu} = - A^{\rm CP}_{\mu e} \equiv A^{\rm CP}_{2\nu} = A^{\rm CPT}_{2\nu}$. It is well known that there is no intrinsic CP violation in the two-flavor case, i.e., there are no physical CP-violating phases. For oscillations in vacuum, both $A^{\rm CPT}_{2\nu}$ and $A^{\rm CP}_{2\nu}$ vanish. However, for oscillations in matter, if the intrinsic CPT symmetry is preserved, we conclude that the matter-induced extrinsic CP and CPT asymmetries are equal to each other in the two-flavor case.
\begin{figure}[!t]
\begin{center}
\includegraphics[width=0.7\textwidth]{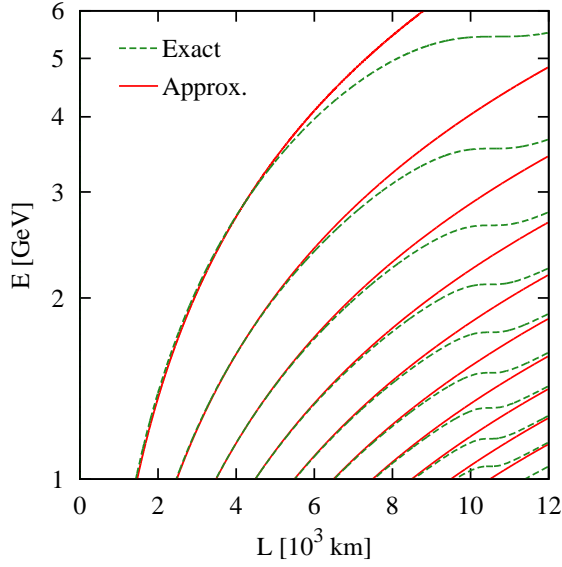}
\vspace{-0.4cm}
\caption{\label{fig:2nu} Contour plot for a vanishing extrinsic CPT asymmetry $A^{\rm CPT}_{2\nu} = 0$, where constant matter density $\rho = 3.5~{\rm g}~{\rm cm}^{-3}$, electron fraction $Y^{}_e = 0.5$, and typical values of the two-flavor neutrino oscillation parameters (i.e., $\sin^2 \theta = 0.024$ and $\Delta m^2 = 2.5\times 10^{-3}~{\rm eV}^2$) are assumed. The solid (red) curves correspond to the approximate and analytical condition $\tan \Delta = 2\Delta$, while the dashed (green) ones to results using the exact two-flavor oscillation probabilities given in Eq.~(\ref{eq:P2nu}).}
\end{center}
\end{figure}

It is straightforward to calculate the extrinsic CPT and CP asymmetries in two-flavor oscillations in matter of constant density. The survival probability is given by \cite{Wolfenstein:1977ue,Mikheev:1986gs}
\begin{eqnarray}
P^{}_{2\nu} = 1 - \frac{\sin^2 2\theta}{r} \sin^2 \left(\frac{\Delta m^2 L}{4E} r \right) \; ,
\label{eq:P2nu}
\end{eqnarray}
where $r = \sqrt{\sin^2 2\theta + \left(\cos 2\theta \pm A \right)^2}$ with $A \equiv 2E V/\Delta m^2$ characterizes the matter effects, and the plus (minus) sign refers to the antineutrino (neutrino) oscillation channel. In the limit of a small matter potential, namely $A \ll 1$, the extrinsic CPT asymmetry turns out to be
\begin{equation}
A^{\rm CPT}_{2\nu} = 2 A \sin^2 2\theta \cos 2\theta \left(2\Delta \cos \Delta - \sin \Delta \right) \sin \Delta + {\cal O}(A^3) \; ,
\label{eq:2nuCPT}
\end{equation}
with $\Delta \equiv \Delta m^2 L/(4E)$. Obviously, the CPT and CP asymmetries are proportional to the matter potential and will vanish for oscillations in vacuum. For a nonzero $A$, the leading-order term in Eq.~(\ref{eq:2nuCPT}) becomes zero if $\tan \Delta = 2\Delta$ is satisfied.

In Fig.~\ref{fig:2nu}, the contour curves for a vanishing CPT asymmetry $A^{\rm CPT}_{2\nu} = 0$ are shown, where constant matter density of $\rho = 3.5~{\rm g}~{\rm cm}^{-3}$, electron fraction $Y^{}_e = 0.5$, and typical neutrino oscillation parameters $\sin^2 \theta = 0.024$ and $\Delta m^2 = 2.5\times 10^{-3}~{\rm eV}^2$ are used. The dashed (green) curves are the exact calculation using neutrino and antineutrino oscillation probabilities given in Eq.~(\ref{eq:P2nu}), while the solid (red) curves correspond to the one using Eq.~(\ref{eq:2nuCPT}) and assuming the leading-order term to vanish, namely, $\tan \Delta = 2\Delta$. It can be observed that the condition $A \ll 1$ is no longer satisfied for a higher neutrino energy and a longer baseline. However, for an extremely-long baseline, the dashed curves obtained from the exact oscillation probabilities approach the solid curves corresponding to two neighboring solutions to $\tan \Delta  = 2 \Delta$.

If the intrinsic CPT symmetry is not preserved, the leptonic mixing angle $\theta$ and mass-squared difference $\Delta m^2$ for neutrinos are generally different from those for antineutrinos. Thus, in order to quantify deviations from the intrinsic CPT symmetry, we express the mixing parameters for antineutrinos as follows
\begin{eqnarray}
\Delta \overline{m}^2 &=& \Delta m^2 (1 + \varepsilon^{}_m) \; , \\
\sin^2 2\overline{\theta} &=& \sin^2 2\theta (1 + \varepsilon^{}_\theta) \; ,
\end{eqnarray}
and expand the CPT asymmetry in terms of perturbation parameters $A$, $\varepsilon^{}_m$, and $\varepsilon^{}_\theta$. Then, we obtain
\begin{eqnarray}
A^{\rm CPT}_{2\nu} \approx A^{\rm CPT}_{2\nu, {\rm ex}} + \varepsilon^{}_m \Delta \sin 2\Delta \sin^2 2\theta + \varepsilon^{}_\theta \sin^2 \Delta \sin^2 2\theta \; ,
\end{eqnarray}
where the first term $A^{\rm CPT}_{2\nu, {\rm ex}}$ refers to the extrinsic CPT asymmetry given in Eq.~(\ref{eq:2nuCPT}). Therefore, an experimental setup for a vanishing or extremely-small extrinsic CPT asymmetry will be sensitive to the intrinsic CPT asymmetry in neutrino oscillations, except for the case where the oscillation terms proportional to the CPT-violating parameters $\varepsilon^{}_m$ and $\varepsilon^{}_\theta$ become extremely small as well.

\subsection{Three-Flavor Case}

Now, we turn to the case of three-flavor oscillations in matter. In general, there are four independent CPT asymmetries, which will be taken as $A^{\rm CPT}_{ee}$, $A^{\rm CPT}_{e \mu}$, $A^{\rm CPT}_{\mu e}$, and $A^{\rm CPT}_{\mu \mu}$ in the following discussion~\cite{Jacobson:2003wc}. For constant matter density, the relevant neutrino oscillation probabilities are given by \cite{Cervera:2000kp,Freund:2001pn,Freund:2001ui,Barger:2001yr,Akhmedov:2004ny}
\begin{eqnarray}
P^{}_{ee} &=& 1 - 4 s^2_{13} \frac{\sin^2 (A-1)\Delta}{(A-1)^2} \; , \label{eq:Pee} \\
P^{}_{e\mu} &=& 4 s^2_{13} s^2_{23} \frac{\sin^2 (A-1)\Delta}{(A-1)^2} + 2 \alpha s^{}_{13} \sin 2\theta_{12} \sin 2\theta_{23} \cos(\Delta - \delta) \frac{\sin A\Delta}{A} \frac{\sin (A-1)\Delta}{A-1} \; , ~~~~~~~~ \label{eq:Pem} \\
P^{}_{\mu\mu} &=& 1 - \sin^2 2\theta_{23} \sin^2 \Delta + \alpha c^2_{12} \sin^2 2\theta_{23} \Delta \sin 2\Delta - 4 s^2_{13} s^2_{23}  \frac{\sin^2 (A-1)\Delta}{(A-1)^2} \nonumber \\
&~& - \frac{2}{A-1} s^2_{13} \sin^2 2\theta_{23} \left[\sin \Delta \cos A\Delta \frac{\sin(A-1)\Delta}{A-1} - \frac{A}{2}\Delta \sin 2\Delta\right] \label{eq:Pmm}
\end{eqnarray}
to second order in $s^{}_{13}$ and first order in $\alpha \equiv \Delta m^2_{21}/\Delta m^2_{31}$. Here $s_{ij} \equiv \sin \theta_{ij}$ and $c_{ij} \equiv \cos \theta_{ij}$ for $ij = 12, 13, 23$ have been defined. In addition, we have defined the oscillation phase driven by the large neutrino mass-squared difference $\Delta m^2_{31}$ as $\Delta \equiv \Delta m^2_{31} L/(4E)$, and $A \equiv 2E V/\Delta m^2_{31}$ that measures the importance of matter effects.  Given current neutrino oscillation data, we have $\alpha \approx \sqrt{2} s^2_{13} \sim 0.03$, so it is safe to neglect ${\cal O}(\alpha^2)$ terms. Note that the series expansion of the oscillation probabilities in Eqs.~(\ref{eq:Pee})--(\ref{eq:Pmm}) is valid as long as $\alpha \Delta \ll 1$, or equivalently, $L/E \ll 10^4~{\rm km}/{\rm GeV}$. Under this condition, the oscillation terms mainly driven by the small mass-squared difference $\Delta m^2_{21}$ are negligible. One can verify that this condition is satisfied by the ongoing and forthcoming long-baseline neutrino oscillation experiments, which make use of intensive neutrino beams of energies around a few GeV and baselines shorter than the diameter of the Earth.

Using constant matter density, it is possible to derive the oscillation probabilities for antineutrinos $\overline{P}_{\alpha \beta}$ from those for neutrinos $P_{\alpha \beta}$ by flipping the signs of the matter potential $V$ (i.e., $A \to -A$) and the CP-violating phase $\delta$ (i.e., $\delta \to -\delta$). Furthermore, the probabilities for the T-conjugate channels $P_{\beta \alpha}$ can be obtained by changing the sign of $\delta$, if the matter density profile is symmetric \cite{Akhmedov:2001kd}, which is obviously the case for constant matter density. Therefore, one can calculate the oscillation probabilities $P_{\mu e}$, $\overline{P}_{ee}$, $\overline{P}_{\mu e}$, $\overline{P}_{e \mu}$, and $\overline{P}_{\mu \mu}$ from  Eqs.~(\ref{eq:Pee})--(\ref{eq:Pmm}) by applying the aforementioned rules. Then, with all the relevant oscillation probabilities, we readily compute the four independent CPT asymmetries
\begin{eqnarray}
A^{\rm CPT}_{ee} &=& 4s^2_{13} \left[\frac{\sin (A+1)\Delta}{A+1} + \frac{\sin (A-1)\Delta}{A-1}\right] \left[\frac{\sin (A+1)\Delta}{A+1} - \frac{\sin (A-1)\Delta}{A-1}\right] \; , \label{eq:Aee} \\
A^{\rm CPT}_{e\mu} &=& -\left\{2\alpha s^{}_{13} \sin 2\theta^{}_{12} \sin 2\theta^{}_{23} \cos(\Delta - \delta) \frac{\sin A\Delta}{A} + 4s^2_{13} s^2_{23}\left[\frac{\sin (A+1)\Delta}{A+1} \right. \right. \nonumber \\ &~& ~~ \left. \left. + \frac{\sin (A-1)\Delta}{A-1}\right]\right\}
\left[\frac{\sin (A+1)\Delta}{A+1} - \frac{\sin (A-1)\Delta}{A-1}\right] \; , \label{eq:Aem} \\
A^{\rm CPT}_{\mu e} &=& -\left\{2\alpha s^{}_{13} \sin 2\theta^{}_{12} \sin 2\theta^{}_{23} \cos(\Delta + \delta) \frac{\sin A\Delta}{A} + 4s^2_{13} s^2_{23}\left[\frac{\sin (A+1)\Delta}{A+1} \right. \right. \nonumber \\ &~& ~~ \left. \left. + \frac{\sin (A-1)\Delta}{A-1}\right]\right\}
\left[\frac{\sin (A+1)\Delta}{A+1} - \frac{\sin (A-1)\Delta}{A-1}\right] \; , \label{eq:Ame} \\
A^{\rm CPT}_{\mu \mu} &=& \left\{2\alpha s^{}_{13} \sin 2\theta^{}_{12} \sin 2\theta^{}_{23} \cos \Delta  \cos \delta \frac{\sin A\Delta}{A} + 4s^2_{13} s^2_{23}\left[\frac{\sin (A+1)\Delta}{A+1} \right. \right. \nonumber \\ &~& ~~ \left. \left. + \frac{\sin (A-1)\Delta}{A-1}\right]\right\}
\left[\frac{\sin (A+1)\Delta}{A+1} - \frac{\sin (A-1)\Delta}{A-1}\right] - \frac{2A\Delta}{A^2 - 1} \sin 2\Delta \nonumber \\
&~& + 2s^2_{13} \sin^2 2\theta_{23} \left[\frac{\sin (A+1)\Delta}{(A+1)^2} + \frac{\sin (A-1)\Delta}{(A-1)^2}\right] \; , \label{eq:Amm}
\end{eqnarray}
where the higher-order terms of ${\cal O}(\alpha s_{13})$ and ${\cal O}(\alpha^2)$ have been neglected. Although it is impossible to obtain a universal condition for all four CPT asymmetries to vanish, one can easily figure out if the following identity
\begin{eqnarray}
\frac{\sin (A+1)\Delta}{A+1} - \frac{\sin (A-1)\Delta}{A-1} = 0 \label{eq:cond1}
\end{eqnarray}
is fulfilled, $A^{\rm CPT}_{ee} = A^{\rm CPT}_{e\mu} = A^{\rm CPT}_{\mu e} = 0$ holds at leading order. This equality is trivially satisfied for $A = 0$, i.e., for oscillations in vacuum. However, there exist non-trivial solutions to Eq.~(\ref{eq:cond1}), as we will show later. The asymmetry $A^{\rm CPT}_{\mu \mu}$ is generally nonzero under this condition, but it can be further reduced to
\begin{eqnarray}
A^{\rm CPT}_{\mu \mu} = \frac{2A}{A^2-1} \left[2s^2_{13} \sin^2 2\theta_{23} \frac{\sin (A-1)\Delta}{A-1} - \sin \Delta\right] \; ,
\label{eq:Ammred}
\end{eqnarray}
which is proportional to $A$ and becomes extremely small for low neutrino energies. Moreover, one can observe that $A^{\rm CPT}_{ee} = 0$ holds if another condition
\begin{eqnarray}
\frac{\sin (A+1)\Delta}{A+1} + \frac{\sin (A-1)\Delta}{A-1} = 0 \label{eq:cond2}
\end{eqnarray}
is satisfied. In this case, we expect $A^{\rm CPT}_{e\mu}$, $A^{\rm CPT}_{\mu e}$, and $A^{\rm CPT}_{\mu \mu}$ to be suppressed as well, since the terms of ${\cal O}(s^2_{13})$ in the first lines of Eqs.~(\ref{eq:Aem})--(\ref{eq:Amm}) vanish and the much smaller terms of ${\cal O}(\alpha s_{13})$ survive.
\begin{figure}[!t]
\begin{center}
\subfigure{%
\hspace{-1.2cm}
\includegraphics[width=0.7\textwidth]{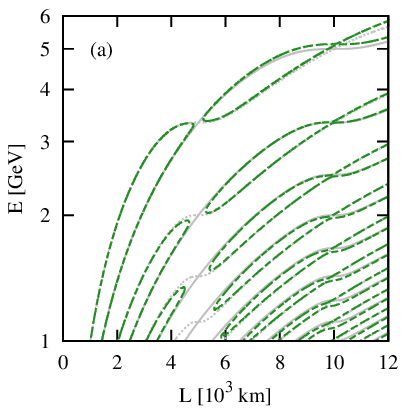}        }%
\subfigure{%
\hspace{-4cm}
\includegraphics[width=0.7\textwidth]{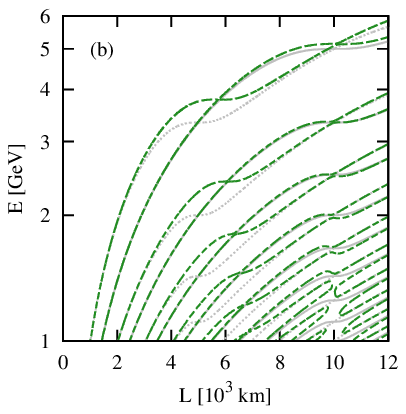}        } \\
\vspace{-0.5cm}
\subfigure{%
\hspace{-1.2cm}
\includegraphics[width=0.7\textwidth]{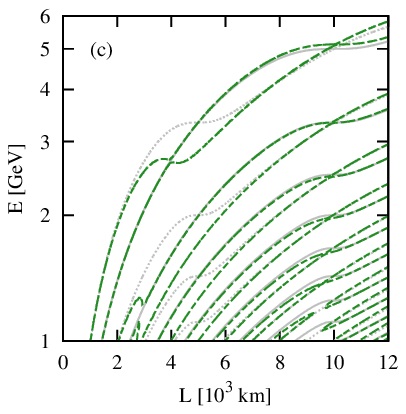}        }%
\subfigure{%
\hspace{-4.0cm}
\includegraphics[width=0.7\textwidth]{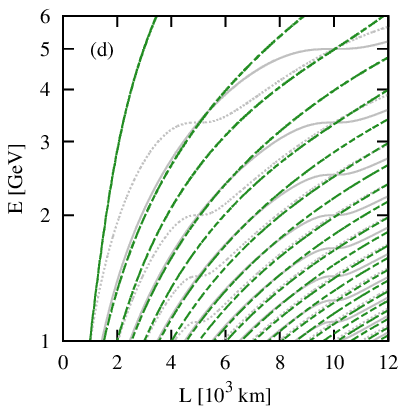}        }
\end{center}
\vspace{-0.8cm}
\caption{\label{fig:magic} Contour plots for vanishing extrinsic CPT asymmetries (a) $A^{\rm CPT}_{ee} = 0$; (b) $A^{\rm CPT}_{e\mu} = 0$; (c) $A^{\rm CPT}_{\mu e} = 0$; (d) $A^{\rm CPT}_{\mu \mu} = 0$, where constant matter density $\rho = 3.5~{\rm g}~{\rm cm}^{-3}$, electron fraction $Y^{}_e = 0.5$, and typical values of the neutrino parameters $\sin^2 \theta_{12} = 0.33$, $\sin^2 \theta_{23} = 0.50$, $\sin^2 \theta_{13} = 0.024$, $\delta = \pi/2$, $\Delta m^2_{21} = 7.5\times 10^{-5}~{\rm eV}^2$, and $\Delta m^2_{31} = 2.5\times 10^{-3}~{\rm eV}^2$ are assumed. The solid (gray) curves correspond to $\sin [(A+1)\Delta]/ (A + 1) - \sin[(A-1)\Delta]/(A-1) = 0$, while the dotted (gray) ones to $\sin [(A+1)\Delta]/ (A + 1) + \sin[(A-1)\Delta]/(A-1) = 0$. The dashed (green) curves denote the numerical results by using exact three-flavor oscillation probabilities.}
\end{figure}

In order to illustrate the above observations, we plot the contour lines for vanishing CPT asymmetries for a variety of neutrino energies and baseline lengths, as shown in Fig.~\ref{fig:magic}. In our calculations, constant matter density $\rho = 3.5~{\rm g}~{\rm cm}^{-3}$, electron fraction $Y_e = 0.5$, and the neutrino parameters $\sin^2 \theta_{12} = 0.33$, $\sin^2 \theta_{23} = 0.50$, $\sin^2 \theta_{13} = 0.024$, $\delta = \pi/2$, $\Delta m^2_{21} = 7.5\times 10^{-5}~{\rm eV}^2$, and $\Delta m^2_{31} = 2.5\times 10^{-3}~{\rm eV}^2$ are assumed. The dashed (green) curves are determined using the exact probabilities for three-flavor oscillations in matter. The solid (gray) curves correspond to the identity in Eq.~(\ref{eq:cond1}), while the dotted (gray) ones to that in Eq.~(\ref{eq:cond2}).

As one can observe from Fig.~\ref{fig:magic} (a), the main structure of the contours for $A^{\rm CPT}_{ee} = 0$ can be perfectly described by the two analytical conditions in Eqs.~(\ref{eq:cond1}) and (\ref{eq:cond2}), i.e., the solid and dotted curves. For each pair of two curves, there are two intersecting points, one of which is along the line of $L_1 = 5000~{\rm km}$ and the other $L_2 = 10^4~{\rm km}$. For neutrino energies around a few GeV, we find that Eqs.~(\ref{eq:cond1}) and (\ref{eq:cond2}) are equivalent to (i) $\cos A\Delta \sin \Delta = 0$ and (ii) $\sin A\Delta \cos \Delta = 0$. On the other hand, we have
\begin{eqnarray}
\Delta &\equiv& \frac{\Delta m^2_{31} L}{4 E} \approx \pi
\left(\frac{\Delta m^2_{31}}{2.5 \times 10^{-3}~{\rm eV}^2}\right)
\left(\frac{L}{1000~{\rm km}}\right) \left(\frac{1~{\rm
GeV}}{E}\right) \;, \nonumber \\
A &\equiv& \frac{2EV}{\Delta m^2_{31}} \approx 0.1
\left(\frac{E}{1~{\rm GeV}}\right) \left(\frac{2.5 \times 10^{-3}~{\rm
eV}^2}{\Delta m^2_{31}}\right) \left(\frac{Y^{}_e}{0.5}\right)
\left(\frac{\rho}{3.5~{\rm g~cm^{-3}}}\right) \; . \label{eq:ADel}
\end{eqnarray}
Therefore, for the given matter density and electron fraction, $A\Delta \approx 0.1\pi (L/10^3~{\rm km})$, implying $A\Delta = \pi/2$ for $L_1 = 5000~{\rm km}$ and $A\Delta = \pi$ for $L_2 = 10^4~{\rm km}$. Since the baseline length cannot exceed the diameter of the Earth, only these two possibilities are allowed. In the first case with $A\Delta = \pi/2$ and $L_1 = 5000~{\rm km}$, we can further fix neutrino energies at the intersecting points by requiring $\cos \Delta = 0$, or equivalently, $\Delta = (2k+1)\pi/2$, where $k$ is a nonnegative integer. With the help of Eq.~(\ref{eq:ADel}), we obtain $E = 10/(2k+1)~{\rm GeV}$, leading to $E = 3.3~{\rm GeV}$, $2.0~{\rm GeV}$, $1.4~{\rm GeV}$, and $1.1~{\rm GeV}$ for $k = 1, 2, 3, 4$, respectively. In the second case with $A\Delta = \pi$ and $L_2 = 10^4~{\rm km}$, the neutrino energies at the intersecting points are further determined by $\sin \Delta = 0$, or equivalently, $\Delta = k\pi$ with $k$ being a nonnegative integer. In a similar way, one can figure out the energies $E = 10/k~{\rm GeV}$ by setting $k = 2, \ldots, 10$. However, it is worthwhile to point out that the oscillation probabilities themselves are also highly suppressed at these points, rendering them not useful in searching for intrinsic CPT violation.

In Fig.~\ref{fig:magic} (b) and (c), the solid curves from the condition in Eq.~(\ref{eq:cond1}) coincide with the dashed curves from the exact numerical calculations. However, the dotted curves from the condition in Eq.~(\ref{eq:cond2}) significantly deviate from the dashed ones. In addition, the analytical conditions in Eqs.~(\ref{eq:cond1}) and (\ref{eq:cond2}) cannot provide a satisfactory description of $A^{\rm CPT}_{\mu \mu} = 0$, as shown in Fig.~\ref{fig:magic} (d). For a baseline length below $5000~{\rm km}$, we have verified that the numerical results with constant matter density in Fig.~\ref{fig:magic} are essentially unchanged when a realistic density profile (e.g., the Preliminary Reference Earth Model~\cite{Dziewonski:1981xy}) is used.

In analogy to the case of two-flavor oscillations, one can introduce different mixing parameters for antineutrinos and  investigate the CPT asymmetries $A^{\rm CPT}_{\alpha \beta}$ in the presence of intrinsic CPT violation. However, with six additional mixing parameters for antineutrinos, the approximate and analytical expressions of $A^{\rm CPT}_{\alpha \beta}$ will be rather lengthy and less instructive. In the next section, we will summarize the current experimental constraints on the antineutrino parameters $(\overline{\theta}_{12}, \overline{\theta}_{13}, \overline{\theta}_{23})$, $(\Delta \overline{m}^2_{21}, \Delta \overline{m}^2_{31})$, and $\overline{\delta}$, whose deviations from the neutrino parameters are clear signatures of intrinsic CPT violation. Moreover, we focus on a future low-energy neutrino factory and a super-bean experiment, and study their sensitivities to the difference between the CP-violating phase $\delta$ in the neutrino sector and $\overline{\delta}$ in the antineutrino sector.

\section{Experimental Constraints}
\label{sec:constr}

If the fundamental CPT symmetry is not assumed, one has to fit neutrino and antineutrino oscillation experiments separately using different mixing parameters and mass-squared differences. In this section, we present a brief summary of current experimental constraints, and emphasize that the future neutrino facilities offer a new possibility to constrain the difference between neutrino and antineutrino CP-violating phases.

\subsection{Current Constraints}

First, we consider the most precise measurements of $\Delta m^2_{21}$ and $\theta_{12}$ in solar neutrino experiments, and $\Delta \overline{m}^2_{21}$ and $\overline{\theta}_{12}$ in the long-baseline reactor neutrino experiment, i.e., KamLAND. In Ref.~\cite{Aharmim:2011vm}, a combined analysis of three phases of solar neutrino data from the SNO experiment has been performed. If the solar neutrino rates in Gallium~\cite{Abdurashitov:2009tn} and Chlorine~\cite{Cleveland:1998nv} experiments, Borexino~\cite{Bellini:2008mr,Bellini:2011rx} and Super-Kamiokande~\cite{Hosaka:2005um,Cravens:2008aa,Abe:2010hy} solar data are further included, a global analysis in the framework of three-flavor oscillations yields~\cite{Aharmim:2011vm}
\begin{equation}
\tan^2 \theta_{12} = 0.436^{+0.048}_{-0.036} \;, ~~~~ \Delta m^2_{21} = \left(5.13^{+1.49}_{-0.98}\right) \times 10^{-5}~{\rm eV}^2 \; ,
\label{eq:th12dm21nu}
\end{equation}
where the $\pm 1\sigma$ errors are attached to the best-fit values, and $\sin^2 \theta_{13} < 0.058$ at $95\%$ confidence level (C.L.). Given $\Delta m^2_{21}$ in the currently-favored region, the $^8{\rm B}$ neutrinos with relatively high energies experience adiabatic flavor conversion in solar matter and the survival probability is just determined by the mixing angle $\theta_{12}$. Thus, the ratio of charged-current neutrino events and the neutral-current ones from the SNO experiment will be very sensitive to the mixing angle $\theta_{12}$, but not the mass-squared difference $\Delta m^2_{21}$. For solar neutrinos of lower energies, matter effects are negligible and the vacuum oscillation probability averaged over the long distance between the Sun and the Earth is applicable. A nonzero $\theta_{13}$ leads to an energy-independent suppression of the survival probability in the three-flavor case, so solar neutrino experiments also place a bound on $\theta_{13}$. The KamLAND experiment is designed to observe the disappearance of $\overline{\nu}_e$ from nuclear reactors at an averaged distance of $180~{\rm km}$, so it is sensitive to $\Delta \overline{m}^2_{21}$ and $\overline{\theta}_{12}$, and also constrains $\overline{\theta}_{13}$. The latest three-flavor analysis of oscillation data in KamLAND indicates~\cite{Gando:2010aa}
\begin{equation}
\tan^2 \overline{\theta}_{12} = 0.436^{+0.102}_{-0.081} \; , ~~~~ \Delta \overline{m}^2_{21} = \left(7.49^{+0.20}_{-0.20}\right) \times 10^{-5}~{\rm eV}^2 \; ,
\label{eq:th12dm21nubar}
\end{equation}
where the best-fit values with $\pm 1\sigma$ errors are given, and $\sin^2 \overline{\theta}_{13} < 0.094$ at $90\%$ C.L. The energy spectrum of neutrino events measured in KamLAND allows us to probe $\Delta \overline{m}^2_{21}$ with a high precision, while the uncertainty in the flux normalization limits the sensitivity to $\overline{\theta}_{12}$.

Note that the bound on $\overline{\theta}_{13}$ from KamLAND should be superseded by the precise measurements from the short-baseline reactor experiments. The determination of $\overline{\theta}_{13}$ is dominated by the Daya Bay experiment, which has recently published the rate~\cite{An:2012eh,An:2013uza} and spectral~\cite{An:2013zwz} measurements of reactor antineutrinos, and an independent measurement via neutron capture on Hydrogen~\cite{An:2014ehw}. The combined analysis of both rate and spectral data from Daya Bay gives
\begin{equation}
\sin^2 2\overline{\theta}_{13} = 0.090^{+0.008}_{-0.009} \; , ~~~~ \Delta \overline{m}^2_{31} = \left(2.59^{+0.19}_{-0.20}\right) \times 10^{-3}~{\rm eV}^2 \; ,
\label{eq:th13dm31nubar}
\end{equation}
where $\Delta \overline{m}^2_{31} \approx \Delta \overline{m}^2_{32}$ is assumed and the tiny difference $\Delta \overline{m}^2_{21}$ is neglected. In addition, we assume normal mass hierarchy in both neutrino and antineutrino sectors throughout this work. The information on $\theta_{13}$ can be extracted from a three-flavor analysis of solar and atmospheric neutrino data, and from the $\nu_\mu \to \nu_e$ appearance data in the accelerator neutrino experiments. The T2K collaboration has carried out a combined analysis of the $\nu_\mu \to \nu_\mu$ disappearance and $\nu_\mu \to \nu_e$ appearance data in the three-flavor oscillation case~\cite{Abe:2013hdq,Abe:2014ugx, dePerio:2014zna}, and obtained $\sin^2 \theta^{}_{23} = 0.520^{+0.045}_{-0.05}$, $\sin^2 \theta^{}_{13} = 0.0454^{+0.011}_{-0.014}$, and $\Delta m^2_{32} = \left(2.51^{+0.11}_{-0.12}\right) \times 10^{-3}~{\rm eV}^2$, where the CP-violating phase $\delta$ is set to be free in the fit.

Then, we come to the measurements of $(\Delta m^2_{32}, \theta^{}_{23})$ and $(\Delta \overline{m}^2_{32}, \overline{\theta}^{}_{23})$ in atmospheric and accelerator neutrino experiments, where both $\nu_\mu \to \nu_\mu$ and $\overline{\nu}_\mu \to \overline{\nu}_\mu$ disappearance channels are dominant. In Ref.~\cite{Abe:2011ph}, a search for differences between the neutrino and the antineutrino oscillation parameters has been performed for all three phases of atmospheric neutrino data in Super-Kamiokande, indicating $-2.6\times 10^{-3}~{\rm eV}^2 < \Delta \overline{m}^2_{32} - \Delta m^2_{32} < 5.3\times 10^{-3}~{\rm eV}^2$ and $-0.25 < \sin^2 2\overline{\theta}_{23} - \sin^2 2\theta_{23} < 0.11$ at $99\%$ C.L. On the other hand, the MINOS experiment has operated in both neutrino and antineutrino channels, and accumulated about 38 kiloton-years of atmospheric neutrinos~\cite{Adamson:2013whj}. The simultaneous fit to neutrino and antineutrino data at MINOS yields~\cite{Cao:2014eca}
\begin{eqnarray}
\sin^2 2\theta^{}_{23} &=& 0.955^{+0.037}_{-0.039} \; , ~~~ \Delta m^2_{32} = \left(2.38^{+0.11}_{-0.90}\right) \times 10^{-3}~{\rm eV}^2 \; , \nonumber \\
\sin^2 2\overline{\theta}^{}_{23} &=& 0.975^{+0.025}_{-0.085} \; , ~~~ \Delta \overline{m}^2_{32} = \left(2.50^{+0.24}_{-0.24}\right) \times 10^{-3}~{\rm eV}^2 \; , \label{eq:th23dm32}
\end{eqnarray}
where the slightly worse sensitivity to antineutrino parameters can be ascribed to a factor of three lower exposure in the accelerator data, and a smaller antineutrino cross section in the atmospheric data.

Finally, by combining Eqs.~(\ref{eq:th12dm21nu})--(\ref{eq:th23dm32}), we summarize the most conservative constraints at $3\sigma$ C.L. from current oscillation data:
\begin{eqnarray}
\left|\Delta m^2_{21} - \Delta \overline{m}^2_{21}\right| &<& 5.9\times 10^{-5}~{\rm eV}^2 \; , \nonumber \\
\left|\Delta m^2_{31} - \Delta \overline{m}^2_{31}\right| &<& 1.1\times 10^{-3}~{\rm eV}^2 \; , \nonumber \\
\left|\sin^2 \theta^{}_{12} - \sin^2 \overline{\theta}^{}_{12}\right| &<& 0.25 \; , \nonumber \\
\left|\sin^2 \theta^{}_{13} - \sin^2 \overline{\theta}^{}_{13}\right| &<& 0.03 \; , \nonumber \\
\left|\sin^2 \theta^{}_{23} - \sin^2 \overline{\theta}^{}_{23}\right| &<& 0.44 \; ,
\end{eqnarray}
and there is essentially no constraint on $|\delta - \overline{\delta}|$ at $3\sigma$ C.L. In deriving the above limits, we have assumed Gaussian errors and chosen the larger absolute value of $\pm 1\sigma$ errors on the relevant oscillation parameters.

In principle, a three-flavor global-fit analysis to all the above neutrino and antineutrino data is needed to derive statistically reliable constraints on the differences between the neutrino and the antineutrino oscillation parameters, which is beyond the scope of our work. Such a analysis was actually performed in Refs.~\cite{GonzalezGarcia:2007ib,Maltoni:2007zf} for the oscillation data at that time, and the expected sensitivities of future beta-beam experiment, medium-baseline reactor experiments, and neutrino factories have been discussed in Ref.~\cite{Antusch:2008zj}. The recent discovery of a nonzero $\overline{\theta}^{}_{13}$ has triggered tremendous discussion in the literature about future experimental sensitivities to the leptonic Dirac CP-violating phase at a low-energy neutrino factory (LENF), which provides the unique possibility to probe the differences in neutrino and antineutrino CP-violating phases as we will show in the next subsection. See Ref.~\cite{Choubey:2011zzq} for a detailed description of different neutrino factory setups.

\subsection{CP-violating Phases}
\begin{figure}[!t]
\begin{center}
\subfigure{%
\hspace{-1.8cm}
\includegraphics[width=0.7\textwidth]{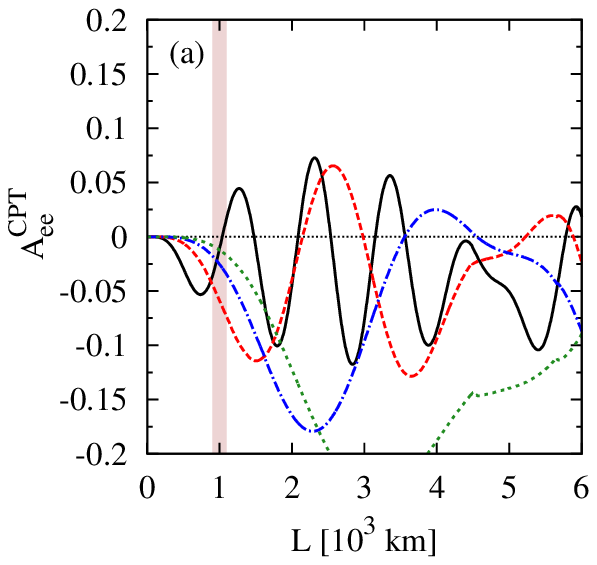}        }%
\subfigure{%
\hspace{-3.2cm}
\includegraphics[width=0.7\textwidth]{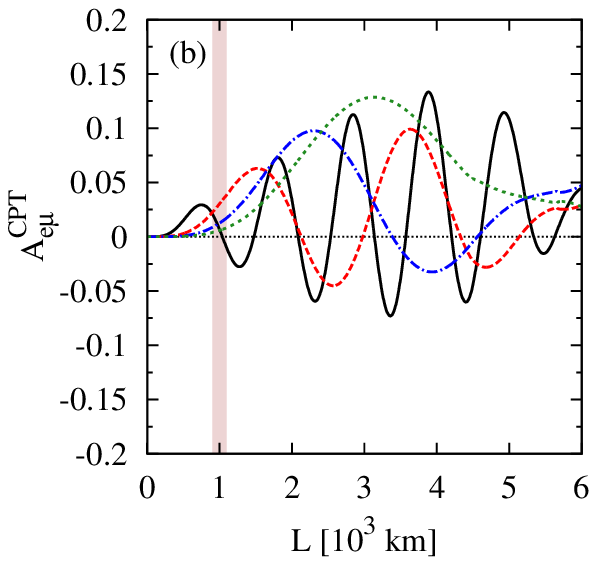}        } \\
\vspace{-0.5cm}
\subfigure{%
\hspace{-1.8cm}
\includegraphics[width=0.7\textwidth]{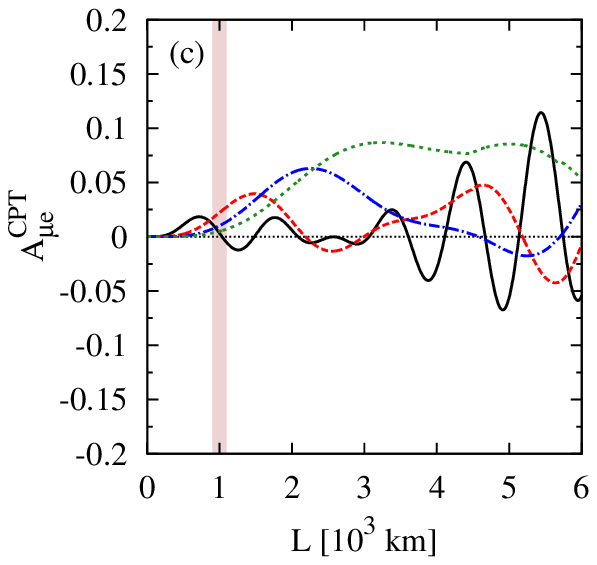}        }%
\subfigure{%
\hspace{-3.2cm}
\includegraphics[width=0.7\textwidth]{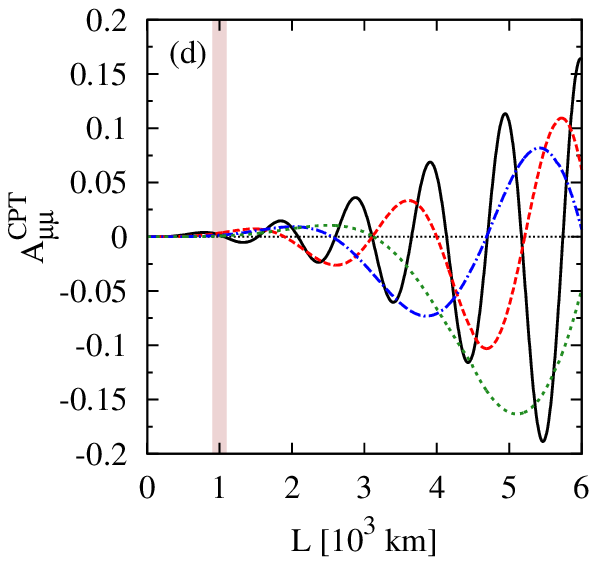}        }
\end{center}
\vspace{-0.8cm}
\caption{\label{fig:cpt} The CPT asymmetries $A^{\rm CPT}_{ee}$, $A^{\rm CPT}_{e\mu}$, $A^{\rm CPT}_{\mu e}$, and $A^{\rm CPT}_{\mu \mu}$ as functions of the baseline length for four different neutrino energies $E = 1~{\rm GeV}$ (black and solid), $2~{\rm GeV}$ (red and dashed), $3~{\rm GeV}$ (blue and dotted-dashed), and $4~{\rm GeV}$ (green and dotted), where the shaded region denotes $L \in [900, 1100]~{\rm km}$. }
\end{figure}
In order to concentrate on the determination of CP-violating phases $\delta$ and $\overline{\delta}$, we assume normal mass hierarchy in both neutrino and antineutrino sectors (i.e., $m^{}_1 < m^{}_2 < m^{}_3$ and $\overline{m}^{}_1 < \overline{m}^{}_2 < \overline{m}^{}_3$). For relatively large $\theta^{}_{13}$ and $\overline{\theta}^{}_{13}$, it has been proposed that a neutrino factory with neutrino energies of several GeV and baseline lengths around $1000~{\rm km}$ will be a powerful facility to pin down the CP-violating phases~\cite{Geer:2007kn,FernandezMartinez:2007ms}. Therefore, we examine the expected sensitivities of a LENF to both $\delta$ and $\overline{\delta}$.

First, it may be instructive to investigate the extrinsic CPT asymmetries at the probability level for a LENF. In Fig.~\ref{fig:cpt}, the four CPT asymmetries $A^{\rm CPT}_{ee}$, $A^{\rm CPT}_{e\mu}$, $A^{\rm CPT}_{\mu e}$, and $A^{\rm CPT}_{\mu \mu}$ for different baseline lengths and neutrino energies are shown. In the numerical calculations, the averaged matter density along the trajectory is used and the full three-flavor oscillation probabilities are implemented. In addition, the neutrino oscillation parameters $\sin^2 \theta_{12} = 0.307$, $\sin^2 \theta_{23} = 0.448$, $\sin^2 \theta_{13} = 0.0242$, $\delta = \pi/2$, $\Delta m^2_{21} = 7.54\times 10^{-5}~{\rm eV}^2$, and $\Delta m^2_{31} = 2.43\times 10^{-3}~{\rm eV}^2$ have been assumed for both neutrinos and antineutrinos. Two comments on the numerical results in Fig.~\ref{fig:cpt} are in order:
\begin{enumerate}
\item As expected, the extrinsic CPT asymmetries are absent in the limit of a very short baseline, when matter effects are negligible. The conditions for vanishing CPT asymmetries, which have been discussed in the previous section, cannot be satisfied for a single baseline length and a wide range of neutrino energies. However, for a LENF with the stored muon energy $E^{}_\mu = 4.5~{\rm GeV}$, the CPT asymmetries are small around $L = 1000~{\rm km}$. One can observe from Fig.~\ref{fig:cpt} that the zero point of CPT asymmetries for the neutrino energy of $E = 1~{\rm GeV}$ (solid black curves) is reached around $L = 1000~{\rm km}$, while the asymmetries for higher neutrino energies have not yet developed much at this baseline length.

\item In Fig.~\ref{fig:cpt} (d), it is evident that $A^{\rm CPT}_{\mu \mu}$ is extremely small for the whole relevant energy range and its absolute value is less than $2.5~\%$ up to the baseline length $L = 2500~{\rm km}$. This has already been observed in Refs~\cite{Barger:2000iv,Xing:2001ys,Bilenky:2001ka}, and it has been proposed that the $\nu^{}_\mu \to \nu^{}_\mu$ disappearance channel is suitable to probe intrinsic CPT violation ~\cite{Barger:2000iv,Bilenky:2001ka}, namely the differences between $(\Delta m^2_{31}, \sin^2 \theta^{}_{23})$ and $(\Delta \overline{m}^2_{31}, \sin^2 \overline{\theta}^{}_{23})$. Due to $A^{\rm T}_{\mu \mu} = 0$, we have $A^{\rm CPT}_{\mu \mu} = A^{\rm CP}_{\mu \mu}$, as a consequence of the fact that sizable differences between neutrino and antineutrino oscillation probabilities appear only at long baselines.
\end{enumerate}
However, it is difficult to conclude from the CPT asymmetries at the probability level that $L = 1000~{\rm km}$ is the optimal baseline length to probe intrinsic CPT violation, since a shorter baseline means a larger number of neutrino events.

Then, we use the GLoBES software~\cite{Huber:2004ka,Huber:2007ji} to perform numerical simulations to study the experimental sensitivity to intrinsic CPT violation at a LENF, in particular to the difference between $\delta$ and $\overline{\delta}$. In the simulation, the baseline length is $1300~{\rm km}$, namely the distance between Fermilab to the Sanford Underground Laboratory at Homestake, South Dakota, USA. This choice is also motivated by our previous observations on the extrinsic CPT asymmetries. Following Ref.~\cite{FernandezMartinez:2010zza}, for the neutrino beam, we assume a muon energy of $4.5~{\rm GeV}$ with $1.4\times 10^{21}$ useful muon decays per year, running for ten years at each polarity. For the detector, we consider a totally active scintillating detector with a fiducial mass of 20 kiloton, a energy threshold of $0.5~{\rm GeV}$, and a $10~\%$ energy resolution. At a neutrino factory, the combination of $\nu^{}_e \to \nu^{}_\mu$ ($\overline{\nu}^{}_e \to \overline{\nu}^{}_\mu$) channels and $\overline{\nu}^{}_\mu \to \overline{\nu}^{}_e$ ($\nu^{}_\mu \to \nu^{}_e$) channels can solve the problem of parameter degeneracies. Moreover, although the disappearance channels $\nu^{}_e \to \nu^{}_e$ ($\overline{\nu}^{}_e \to \overline{\nu}^{}_e$) and $\overline{\nu}^{}_\mu \to \overline{\nu}^{}_\mu$ ($\nu^{}_\mu \to \nu^{}_\mu$) are insensitive to the CP-violating phase, they are helpful in determining the other mixing parameters. Hence, we include all these signal channels in our simulations. Note that we have explicitly indicated the signals in the case of $\mu^+$ decays, whereas those in the case of $\mu^-$ decays are given in the parentheses.

In the $\nu^{}_\mu$ and $\overline{\nu}^{}_\mu$ (dis)appearance channels, the detection efficiency of $\mu^\pm$ is set to be $73~\%$ below $1~{\rm GeV}$ and $94~\%$ above. The main background arises from the charge misidentification and neutral-current events, for which we assume a constant fraction $0.1~\%$ of the wrong-sign rates and the neutral-current rates. The charge identification of $e^\pm$ in the low-energy region is very challenging, and the pion background is difficult to subtract from the electron signals. Therefore, in the $\nu^{}_e$ and $\overline{\nu}^{}_e$ (dis)appearance channels, the detection efficiency of $e^\pm$ is set to be $37~\%$ below $1~{\rm GeV}$ and $47~\%$ above. Furthermore, we assume the same type of background as in the $\mu^\pm$ case, and choose a constant fraction of $1~\%$ for the wrong-sign rates and the neutral-current rates. For both cases, an uncorrelated systematic error of $2~\%$ on signal and background is adopted.
\begin{figure}[!t]
\begin{center}
\subfigure{%
\hspace{-1.0cm}
\includegraphics[width=0.6\textwidth]{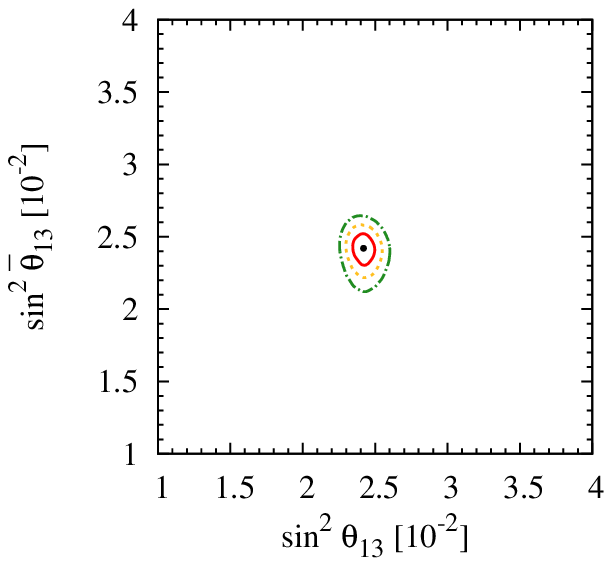}        }%
\subfigure{%
\hspace{-2.5cm}
\includegraphics[width=0.6\textwidth]{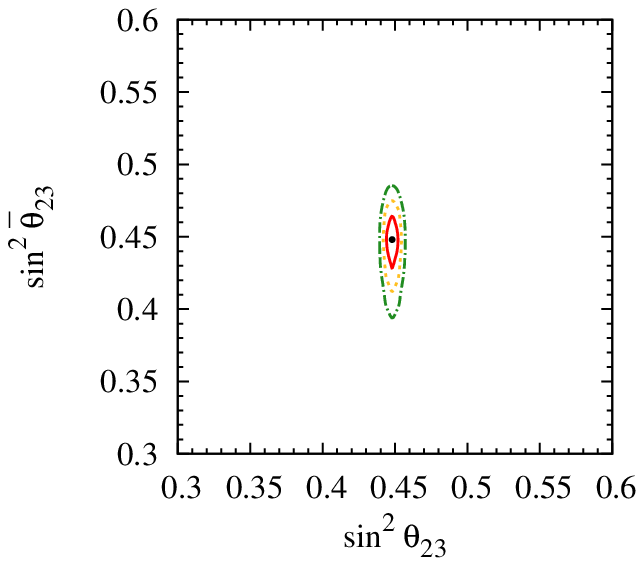}        } \\
\vspace{-0.5cm}
\subfigure{%
\hspace{-1.0cm}
\includegraphics[width=0.6\textwidth]{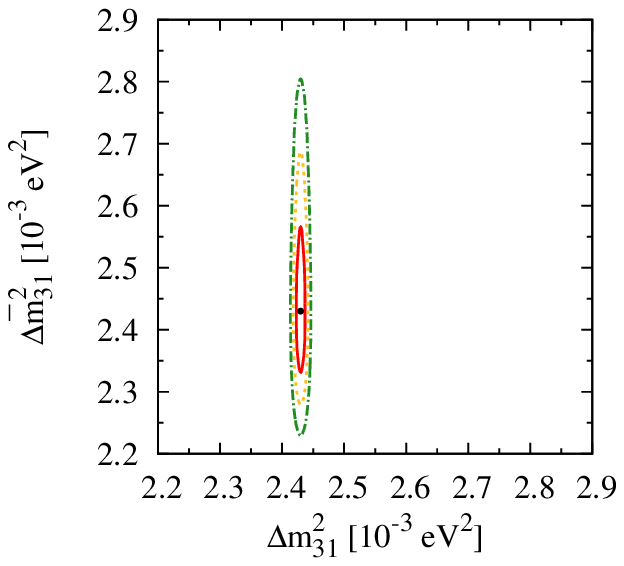}        }%
\subfigure{%
\hspace{-2.5cm}
\includegraphics[width=0.6\textwidth]{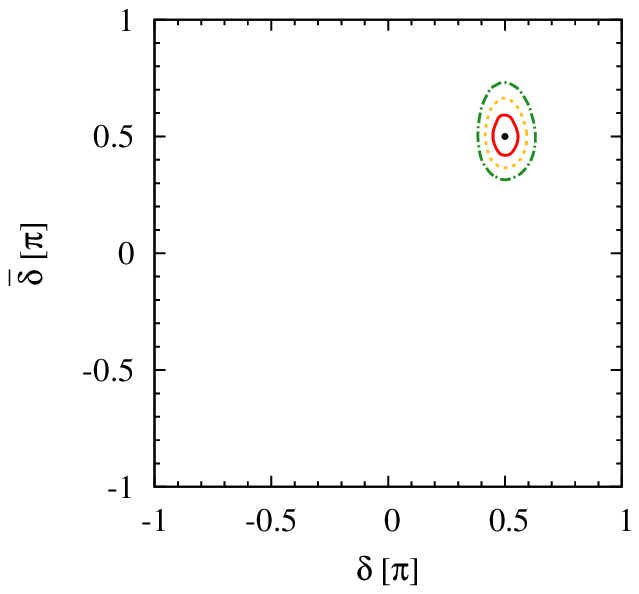}        }
\vspace{-0.5cm}
\end{center}
\vspace{-0.8cm}
\caption{\label{fig:LENF}Future sensitivities to $(\sin^2 \theta_{13}, \sin^2 \overline{\theta}_{13})$, $(\sin^2 \theta_{23}, \sin^2 \overline{\theta}_{23})$, $(\Delta m^2_{31}, \Delta \overline{m}^2_{31})$, and $(\delta, \overline{\delta})$ at a low-energy neutrino factory, where the stored muon energy is $4.5~{\rm GeV}$ and $1.4\times 10^{21}$ useful muon decays per year are assumed for each polarity running for ten years. A $20~{\rm kt}$ totally active scintillator detector is implemented and the baseline length is set to $1300~{\rm km}$.}
\end{figure}

\begin{figure}[!t]
\begin{center}
\subfigure{%
\hspace{-1.0cm}
\includegraphics[width=0.6\textwidth]{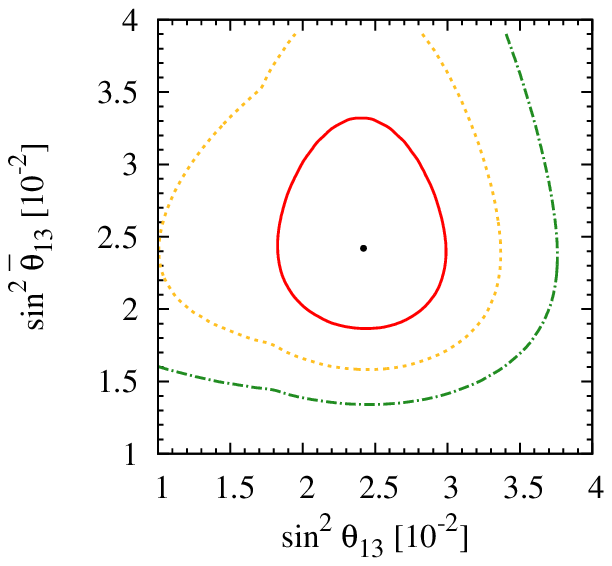}        }%
\subfigure{%
\hspace{-2.5cm}
\includegraphics[width=0.6\textwidth]{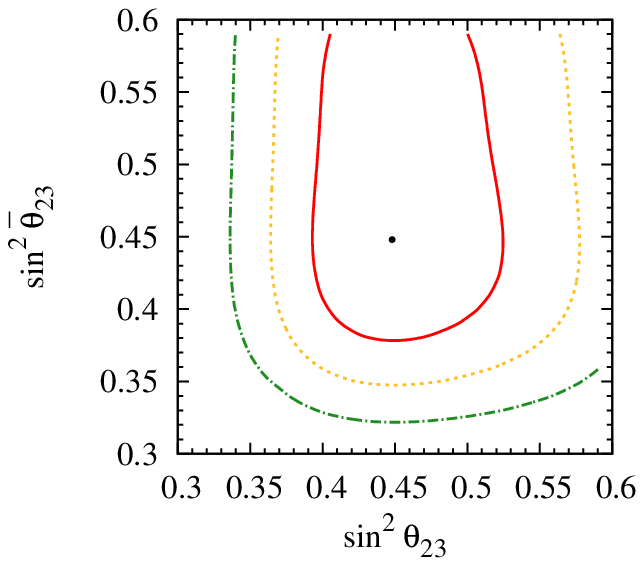}        } \\
\vspace{-0.5cm}
\subfigure{%
\hspace{-1.0cm}
\includegraphics[width=0.6\textwidth]{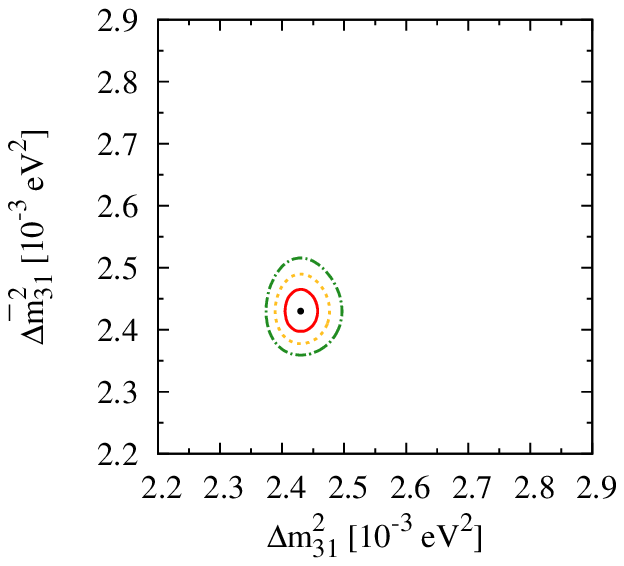}        }%
\subfigure{%
\hspace{-2.5cm}
\includegraphics[width=0.6\textwidth]{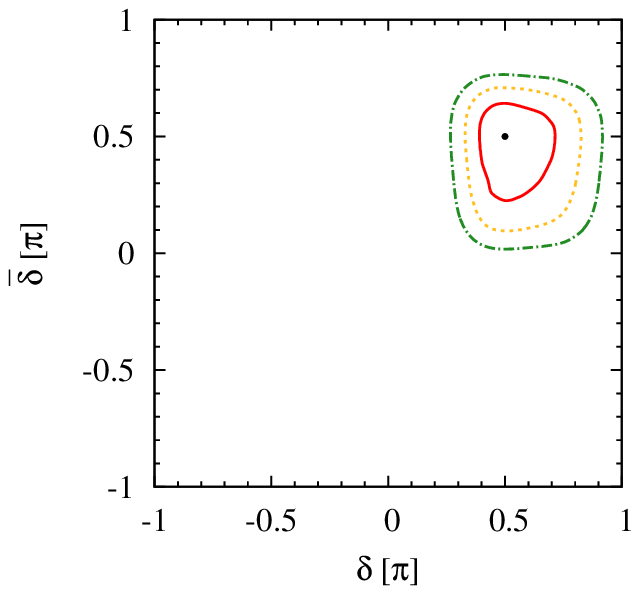}        }
\vspace{-0.5cm}
\end{center}
\vspace{-0.8cm}
\caption{\label{fig:ESSnuSB}Future sensitivities to $(\sin^2 \theta_{13}, \sin^2 \overline{\theta}_{13})$, $(\sin^2 \theta_{23}, \sin^2 \overline{\theta}_{23})$, $(\Delta m^2_{31}, \Delta \overline{m}^2_{31})$, and $(\delta, \overline{\delta})$ at the ESS$\nu$SB, where the proton beam energy is $2.5~{\rm GeV}$ and the protons on target are $2.2\times 10^{23}$ per year. The super-beam experiment is assumed to run for two years in the neutrino mode and eight years in the antineutrino one and a $500~{\rm kt}$ Cherenkov detector (the same as the MEMPHYS detector) is placed at a distance of $540~{\rm km}$.}
\end{figure}

\begin{figure}[!t]
\begin{center}
\subfigure{%
\hspace{-1.0cm}
\includegraphics[width=0.6\textwidth]{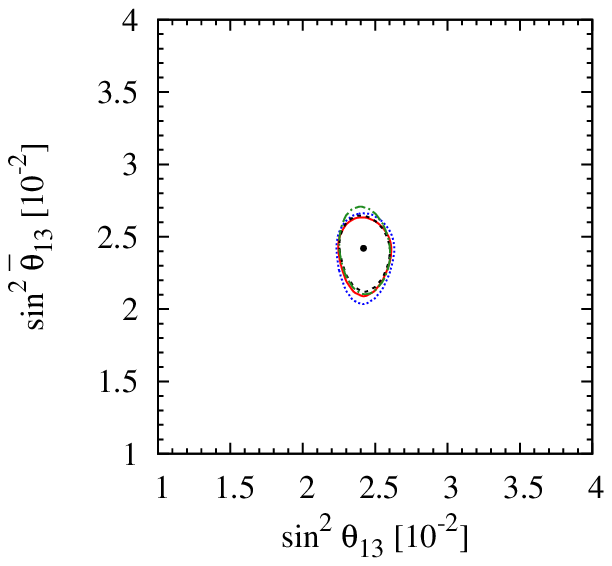}        }%
\subfigure{%
\hspace{-2.5cm}
\includegraphics[width=0.6\textwidth]{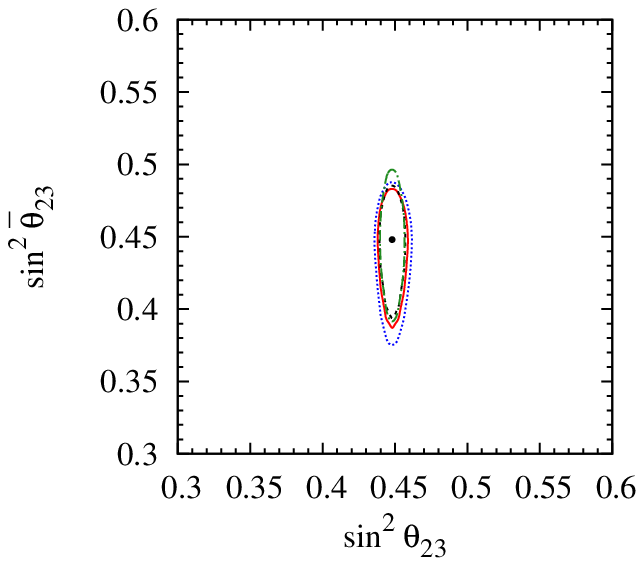}        } \\
\vspace{-0.5cm}
\subfigure{%
\hspace{-1.0cm}
\includegraphics[width=0.6\textwidth]{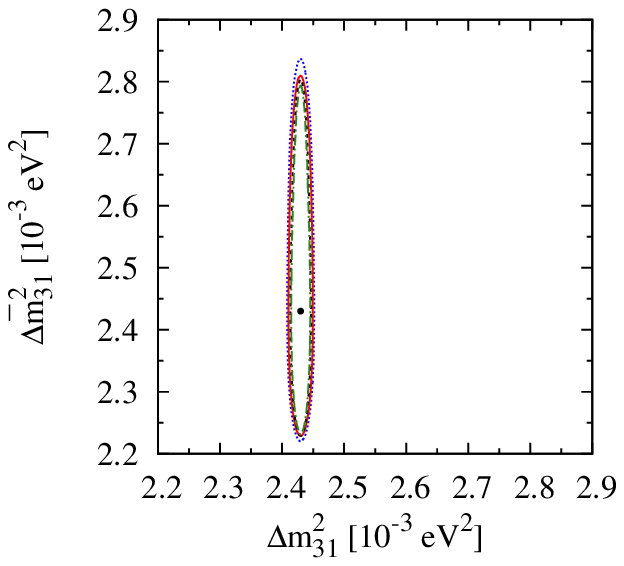}        }%
\subfigure{%
\hspace{-2.5cm}
\includegraphics[width=0.6\textwidth]{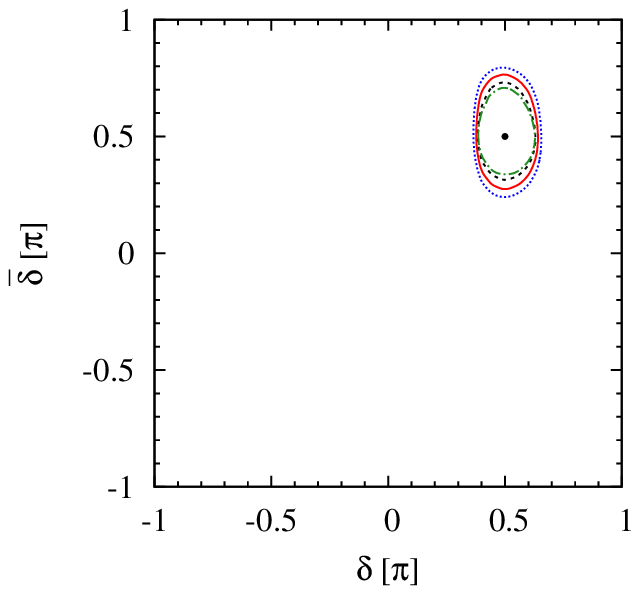}        }
\vspace{-0.5cm}
\end{center}
\vspace{-0.8cm}
\caption{\label{fig:vbaseline}Future sensitivities at $3\sigma$ level to $(\sin^2 \theta_{13}, \sin^2 \overline{\theta}_{13})$, $(\sin^2 \theta_{23}, \sin^2 \overline{\theta}_{23})$, $(\Delta m^2_{31}, \Delta \overline{m}^2_{31})$, and $(\delta, \overline{\delta})$ at a low-energy neutrino factory, where the stored muon energy is $4.5~{\rm GeV}$ and $1.4\times 10^{21}$ useful muon decays per year are assumed for each polarity running for ten years. A $20~{\rm kt}$ totally active scintillator detector is implemented, and the baseline length is set to $800~{\rm km}$ (blue dotted curves), $1000~{\rm km}$ (red solid curves), $1300~{\rm km}$ (black dashed curves), and $1500~{\rm km}$ (green dash-dotted curves).}
\end{figure}

In addition to the LENF, we consider the European Spallation Source Neutrino Super-Beam (ESS$\nu$SB), which has recently been proposed as a promising alternative to probe the leptonic CP-violating phase with a high significance. The nominal setup of ESS$\nu$SB has been described in detail in Ref.~\cite{Baussan:2013zcy}. In the present work, the neutrino fluxes are calculated assuming that the proton beam energy is $2.5~{\rm GeV}$ and the number of protons on target is $2.2\times 10^{23}$ per year. Furthermore, a $500~{\rm kt}$ Cherenkov detector (the same as the MEMPHYS detector) is implemented, so the migration matrices of detector response, selection efficiencies and backgrounds are mainly taken from Ref.~\cite{Agostino:2012fd}.\footnote{The authors are grateful to Enrique Fern{\'a}ndez-Mart{\'i}nez for providing the neutrino fluxes, and to Luca Agostino for the migration matrices of the MEMPHYS detector.} See, Refs.~\cite{Mezzetto:2003mm,Campagne:2004wt,Campagne:2006yx}, for earlier discussions. For illustration, the baseline length is set to be $540~{\rm km}$, which is the distance between the future ESS in Lund, Sweden and the mine in Garpenberg, Sweden. Note that although the performance of a $2.5~{\rm GeV}$ proton energy is generally better than that of the $2.0~{\rm GeV}$ one, given a constant proton power $5~{\rm MW}$, the former setup requires more modifications of the ESS design. A 5~\% (10~\%) systematic error is assumed for the signal (background), which is more optimistic than that considered in Ref.~\cite{Coloma:2012ji,Baussan:2013zcy}.

In our simulations, the oscillation parameters $\sin^2 \theta_{12} = 0.307$, $\sin^2 \theta_{23} = 0.448$, $\sin^2 \theta_{13} = 0.0242$, $\delta = \pi/2$, $\Delta m^2_{21} = 7.54\times 10^{-5}~{\rm eV}^2$, and $\Delta m^2_{31} = 2.43\times 10^{-3}~{\rm eV}^2$ have been assumed for neutrinos, and the same values of the parameters for antineutrinos in the CPT-conserving limit. As observed in Ref.~\cite{Antusch:2008zj}, the experimental sensitivities to $(\sin^2 \theta_{12}, \sin^2 \overline{\theta}_{12})$ and $(\Delta m^2_{21}, \Delta \overline{m}^2_{21})$ can be improved mainly at the medium-baseline reactor and Beta-Beam experiments. Therefore, we focus on the parameters, for which a LENF and ESS$\nu$SB have the best sensitivities, and assume that $(\sin^2 \theta_{12}, \sin^2 \overline{\theta}_{12})$ and $(\Delta m^2_{21}, \Delta \overline{m}^2_{21})$ can be well determined at other future neutrino facilities. Additionally, the current experimental constraints on the relevant oscillation parameters from the previous subsection are taken as priors. The future sensitivities at the LENF and the ESS$\nu$SB facilities are presented in Figs.~\ref{fig:LENF} and \ref{fig:ESSnuSB}, respectively, where we can observe that
\begin{itemize}
\item The LENF is very powerful in constraining the differences between neutrino and antineutrino mixing angles, namely $|\sin^2 \theta^{}_{13} - \sin^2 \overline{\theta}^{}_{13}|$ and $|\sin^2 \theta^{}_{23} - \sin^2 \overline{\theta}^{}_{23}|$, and the CP-violating phases $|\delta - \overline{\delta}|$. The constraint on the antineutrino mass-squared difference $\Delta {\overline{m}}^2_{31}$ is significantly affected by the CP-violating phase $\overline{\delta}$, which has been marginalized over. At this point, the ESS$\nu$SB with a $2\nu + 8\overline{\nu}$ run will do much better. Therefore, the LENF and the ESS$\nu$SB are complementary to each other, and their combination will greatly improve the bounds on those CPT-violating parameters.

\item The ESS$\nu$SB could provide the first possibility to probe the CPT-violating parameter $|\delta - \overline{\delta}|$, which will be more severely constrained by the LENF in the relatively far future. The $3\sigma$ bound reaches $|\delta - \overline{\delta}| \lesssim 0.35\pi$ for the true value $\delta = \overline{\delta} = \pi/2$ in the latter case, whereas it becomes worse by a factor of two, namely $|\delta - \overline{\delta}| \lesssim 0.7\pi$, in the former case. Note that the sensitivity will be much improved at the ultimate neutrino factory, since the neutrino and antineutrino mixing parameters are to be precisely measured in the ongoing and forthcoming oscillation experiments.
\end{itemize}
It is worth stressing that a complete global-fit analysis of current neutrino oscillation data is required to derive the bounds on the intrinsic CPT-violating parameters, whereas a dedicated study of future neutrino facilities taking account of more realistic experimental setups and systematics is necessary to forecast the actual sensitivities. Such an investigation is beyond the scope of the present work, and here we have considered the preliminary version of the LENF and the ESS$\nu$SB for illustration.

In Fig.~\ref{fig:vbaseline}, we examine how the future experimental sensitivity of a low-energy neutrino factory depends on the baseline length. For concreteness, the same experimental setup as in Fig.~\ref{fig:LENF} is used, but the baseline length is assumed to be $L = 800~{\rm km}$, $1000~{\rm km}$, $1300~{\rm km}$ (the same baseline length as used in Fig.~\ref{fig:LENF}), and $1500~{\rm km}$, respectively. Some comments on the results presented in Fig.~\ref{fig:vbaseline} are in order. First, a longer baseline $L = 1500~{\rm km}$ is slightly favored in constraining the difference between $\Delta m^2_{31}$ and $\Delta \overline{m}^2_{31}$, and that between $\delta$ and $\overline{\delta}$, as a consequence of an improved measurement of antineutrino parameters. Second, the default value $L = 1300~{\rm km}$ turns out to be the optimistic one to probe the intrinsic CPT-violating parameters $(\sin^2 \theta_{13}, \sin^2 \overline{\theta}_{13})$ and $(\sin^2 \theta_{23}, \sin^2 \overline{\theta}_{23})$. However, the baseline length $L = 1000~{\rm km}$ has quite a similar performance, which seems reasonable according to the analysis of oscillation probabilities in the previous section. In general, it is difficult to optimize a single experiment for all intrinsic CPT-violating parameters. The feasible way is to measure neutrino and antineutrino oscillation parameters as precisely as possible in a number of different experiments with optimal setups.

\section{Summary and Conclusions}
\label{sec:s&c}

Motivated by the recent remarkable progress in experimental neutrino physics, in parti\-cular the discovery of a sizable mixing angle in the Daya Bay experiment, a lot of attention has been paid to the phenomenological studies of the physics potentials of the planned and proposed neutrino facilities. Among them, the precision measurement of oscillation para\-meters, the determination of the neutrino mass ordering, and the discovery of leptonic CP violation are the most important. Along this line, many interesting proposals of superbeam experiments and neutrino factories are under active discussion.

In the present work, we have reconsidered the extrinsic and intrinsic CPT violation in neutrino oscillation experiments. First, we have presented some general discussions about CP, T, and CPT asymmetries in neutrino oscillations, and pointed out an identity among them, namely, $A^{\rm CP}_{\alpha \beta} = A^{\rm CPT}_{\beta \alpha} + A^{\rm T}_{\alpha \beta}$. Then, a detailed study of the extrinsic CPT viola\-tion, which is caused by Earth matter effects and should be present in any long-baseline neutrino oscillation experiments, has been carried out in both two-flavor and three-flavor cases. In the three-flavor case, we have calculated the CPT asymmetries $A^{\rm CPT}_{ee}$, $A^{\rm CPT}_{e\mu}$, $A^{\rm CPT}_{\mu e}$, and $A^{\rm CPT}_{\mu \mu}$ in the plane of different energies and baseline lengths, and demonstrated that two analytical conditions, i.e., $[\sin (1-A)\Delta]/(1 - A) \pm [\sin(1+A)\Delta]/(1+A) = 0$, can be used to approximately describe the main structure of the contours of vanishing extrinsic CPT asymmetries. Finally, without assuming the fundamental CPT symmetry a priori, we have summarized the current experimental constraints on the differences between neutrino and antineutrino mixing parameters. Furthermore, it has been stressed that superbeam experiments and neutrino factories can probe the difference between the neutrino and antineutrino CP-violating phases, namely $|\delta - \overline{\delta}|$. For illustration, a low-energy neutrino factory and the ESS$\nu$SB have been considered, for which the bound at the $3\sigma$ level has been found to be $|\delta - \overline{\delta}| \lesssim 0.35\pi$ and $|\delta - \overline{\delta}| \lesssim 0.7\pi$, respectively.

With more precise measurements of neutrino and antineutrino mixing parameters in the ongoing and forthcoming oscillation experiments, we will be able to test the standard picture of neutrino oscillations, and even to probe new physics scenarios, such as non-standard neutrino interactions and sterile neutrinos. More importantly, future neutrino facilities will allow us to learn about whether the fundamental CPT symmetry is exactly valid in the neutrino sector or not. The discovery of fundamental CPT violation obviously points to new physics beyond the standard model of elementary particles.

\begin{acknowledgments}
We would like to thank Luca Agostino, Mattias Blennow, Sandhya Choubey, and Enrique Fern{\'a}ndez-Mart{\'i}nez for useful discussions. This work was supported by the Swedish Research Council (Vetenskapsr{\aa}det), contract no.~621-2011-3985.
\end{acknowledgments}

\bibliography{references}

\end{document}